\documentclass[review]{elsarticle}

\usepackage{hyperref}

\usepackage{graphicx}
\usepackage{amsmath}
\usepackage{enumerate}
\usepackage{booktabs}
\usepackage{multirow}
\usepackage{hhline}
\usepackage{amssymb}
\usepackage{color}

\definecolor{OliveGreen}{rgb}{0,0.6,0}

\journal{Journal of \LaTeX\ Templates}

\begin{document}

\begin{frontmatter}

\title{Fully Automated 3D Segmentation of MR-Imaged Calf Muscle Compartments: Neighborhood Relationship Enhanced Fully Convolutional Network}




\author[mymainaddress]{Zhihui Guo\fnref{myfootnote}}
\author[mymainaddress]{Honghai Zhang}
\author[mymainaddress]{Zhi Chen}
\author[mysecondaryaddress]{Ellen van der Plas}
\author[mythird]{Laurie Gutmann}
\author[mysecondaryaddress]{Daniel Thedens}
\author[mysecondaryaddress]{Peggy Nopoulos}
\author[mymainaddress]{Milan Sonka}
\fntext[myfootnote]{Corresponding author. zhihui-guo@uiowa.edu}
\address[mymainaddress]{Iowa Institute for Biomedical Imaging, University of Iowa, Iowa City, IA 52242, USA}
\address[mysecondaryaddress]{Department of Psychiatry, University of Iowa, Iowa City, IA 52242, USA}
\address[mythird]{Department of Neurology, University of Iowa, Iowa City, IA 52242, USA.}

\begin{abstract}
Automated segmentation of individual calf muscle compartments from 3D magnetic resonance (MR) images is essential for developing quantitative biomarkers for muscular disease progression and its prediction. Achieving clinically acceptable results is a challenging task due to large variations in muscle shape and MR appearance. In this paper, we present a novel fully convolutional network (FCN) that utilizes contextual information in a large neighborhood and embeds edge-aware constraints for individual calf muscle compartment segmentations. An encoder-decoder architecture is used to systematically enlarge convolution receptive field and preserve information at all resolutions. Edge positions derived from the FCN output muscle probability maps are explicitly regularized using kernel-based edge detection in an end-to-end optimization framework. Our method was evaluated on 40 T1-weighted MR images of 10 healthy and 30 diseased subjects by 4-fold cross-validation. Mean DICE coefficients of 88.00\%--91.29\% and mean absolute surface positioning errors of 1.04--1.66 mm were achieved for the five 3D muscle compartments.
\end{abstract}


\end{frontmatter}


\section{Introduction}



\begin{figure}[t]
\centering
\includegraphics[width=1.0\linewidth]{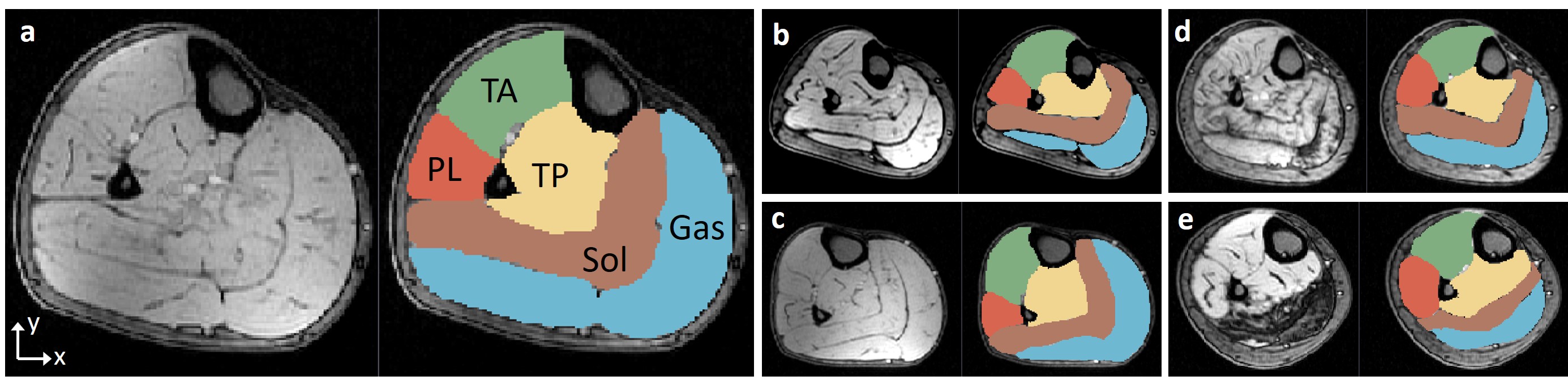}
\caption{Examples of T1-weighted MR images of calf muscle cross sections. Each panel shows the original image and the corresponding expert segmentations for TA, TP, Sol, Gas and PL. (a--c) Normal subjects. (d--e) Patients with severe DM1. Best viewed in color.
} \label{fig:variation}
\end{figure}

Calf muscle is a skeletal muscle group in the lower leg between the knee joint and the ankle, primarily supporting weight-bearing activities such as walking, running, and jumping. Anatomically, the group can be divided into five individual muscle compartments: Tibialis Anterior (TA), Tibialis Posterior (TP), Soleus (Sol), Gastrocnemius (Gas), and Peroneus Longus (PL) \cite{yaman2013magnetic}, as shown in Fig.\ \ref{fig:variation}(a). Volumetric and structural changes of the muscle are important for evaluating muscular disease severity and progression. 
For example, myotonic dystrophy type 1 (DM1), the most common form of inherited muscular dystrophy in adults, causes severe fatty degeneration of calf muscle in most of the patients \cite{wattjes2010neuromuscular}. Magnetic resonance (MR) imaging has been widely used in the clinic for muscular disease diagnosis and follow-up evaluation due to its high sensitivity to dystrophic changes \cite{wattjes2010neuromuscular,stramare2010mri}. Changes in MR images also correlate with clinical outcome measures potentially serving as imaging biomarkers for clinical research \cite{willcocks2016multicenter}. 
Current analysis approaches invariably include hand-tracing of individual compartments that is time consuming and less than ideal for clinical trials. 
Automated segmentation of calf muscle is therefore essential for developing quantitative biomarkers of muscular disease progression and can contribute to its prediction.

A plethora of methods has been developed to separate calf muscle region, subcutaneous adipose tissue (SAT), and intermuscular adipose tissue (IMAT) from the lower leg tissue region, based on which a muscle fat percentage can be obtained as a measurement of fatty degeneration/infiltration that has been shown to have correlation with disease progression \cite{wren2008three}. For example, Valentinitsch \textit{et al.} \cite{valentinitsch2013automated} applied multi-stage K-means clustering \cite{macqueen1967some} to segment calf muscle, SAT and IMAT. Amer \textit{et al.} \cite{amer2019automatic} first used a fully convolutional network (FCN) to segment the whole muscle mask and then classified healthy muscle and IMAT from the segmented mask by deep convolutional auto-encoder. These whole muscle segmentation approaches rely on the intra-object homogeneity to separate muscle and adipose tissues. Afterward, an overall muscle fat percentage can be obtained. However, specific muscle compartments may be more affected in different neuromuscular diseases (e.g., the posterior compartment shows initial changes in this population \cite{heskamp2019lower}). Assessing the changes in individual muscle compartments may be more sensitive than measuring change in the whole calf. In order to study the disease progression in individual muscle compartments \cite{gourgiotis2007acute,alizai2012comparison,commean2011magnetic}, segmentation of individual muscle compartments is critical. 

Different from the entire muscle-region segmentation, automated segmentation of individual muscle compartments is a challenging task due to the unique characteristics of MR muscle images as shown in Fig.\ \ref{fig:variation}. All non-diseased muscle compartments have similar appearances while the MR bias field exists across the whole image region. Besides, muscular dystrophy can introduce substantial appearance changes to a part or the whole compartment (Fig.\ \ref{fig:variation}d-e). Therefore, identifying individual compartments using only local characteristics is unrealistic. On the other hand, shape model based approaches are  unsuitable due to large shape variations and deformations caused by the disease as well as by patient's positioning in the scanner.

Attempts to segment individual muscle compartments on MR images are rare. Essafi \textit{et al.} \cite{essafi2009wavelet} used a landmark-based approach with diffusion wavelets to represent shape variations for 3D segmentation of TA and PL and achieved a mean DICE coefficient of 0.55. Wang \textit{et al.} \cite{wang20103d} encoded shape prior by a point distribution model in a higher-order Markov Random Field framework to segment the medial Gas compartment on the same dataset used in \cite{essafi2009wavelet} and obtained an averaged landmark error around 7~mm. Commean \textit{et al.} \cite{commean2011magnetic} presented a semi-automated method to segment five individual muscle compartments by thresholding and edge detection to study MR imaging measurement reproducibility. Troter \textit{et al.} \cite{le2016volume} used a multi-atlas registration approach for individual muscle segmentation in quadriceps femoris and achieved an averaged DICE of 0.87 $\pm$ 0.11 on an MRI dataset of healthy young adults. Rodrigues \textit{et al.} \cite{rodrigues2019segmentation} adopted a two-stage mechanism to segment individual muscle compartments by first identifying all muscle voxels using Adaboost classifier and then registering the muscle mask to a reference atlas for muscle compartment labeling. However, the individual muscle compartment segmentation results were not visually promising and the accuracies were not reported. 

Compared with traditional techniques, FCN has a large model capacity to learn complex representations and enables pixel-to-pixel training, which makes it suitable for this application. To overcome difficulties mentioned above, an FCN with strengthened neighborhood relationship is desired. Many methods have been reported that imposed high-level neighborhood-aware or edge-aware relationships to either refine FCN outputs or directly change FCN internal architectures. Bauer \textit{et al.} \cite{bauer2011fully} adopted a Conditional Random Field strategy to regularize classification results for brain tumor segmentation. Similarly, Guo \textit{et al.} \cite{guo2018deep} applied a topology-wise graph to refine FCN output for pancreatic tumor segmentation. Both Chen \textit{et al.} \cite{chen2017dcan} and Shen \textit{et al.} \cite{shen2017boundary} used a multi-task FCN to predict object region and edge maps simultaneously for histological object segmentation and brain tumor segmentation respectively, with each task regularized by a cross entropy loss term. Recently, Kampffmeyer \textit{et al.} \cite{kampffmeyer2019connnet} proposed a single-task FCN to predict pixel level connectivity maps based on $n$-neighborhood ($n$=4, 8) relationships, and reverted the prediction to segmentation masks using pixel-pair connectivity agreement. 

In this paper, we propose a novel neighborhood relationship-aware FCN based on a variant of 3D UNet \cite{cciccek20163d}, called FilterNet, for automated segmentation of all five calf muscle compartments. We enhance neighborhood relationships in two ways: efficiently enlarge convolution receptive field and explicitly derive object boundaries directly from object prediction maps in an end-to-end training optimization framework. Specifically, by enlarging the convolution receptive field, information in a larger neighborhood is taken into consideration when generating the prediction for each central voxel, increasing the model robustness.
Additionally, we use kernel-based edge detectors on the prediction maps to regularize the voxel-level probability dissimilarity inside a neighborhood region defined by the kernel size. Motivations behind such kernel edge detector-based constraints are 3-fold. First, the sizes of medical datasets are often small, which makes it not favorable to learn edge regularization from scratch. Edge detectors assess pre-defined neighbor relations (often using derivative formulas) and can be regarded as high-level initialization of the edge regularization module. 
Second, kernel edge detectors played an important role in medical image segmentation approaches over the past several decades, and they are compatible with CNN end-to-end training. Third, this mechanism provides flexibility to further fine-tune hyper-parameters of the kernel, by making the parameters of interest trainable.

Compared with previously reported approaches, our work has several contributions: a) we report a fully automated approach for 3D segmentation of five calf muscle compartments simultaneously; b) by considering the similar textures shared by individual muscles, we are able to efficiently impose edge constrains in an end-to-end training manner; c) our method is robust in MR images from both healthy subjects and patients with DM1. 

To the best of our knowledge, this is the first automated approach for five calf muscle compartments segmentation. Methodologically, our work is the first attempt to regularize neighborhood relationships in the form of kernel-based edge detection on prediction maps that allows direct back-propagation.

\section{Methods}

\subsection{Pre-processing}
In the pre-processing step, a bias field correction method described in \cite{tustison2009n4itk} is first applied to reduce image intensity inhomogeneity by estimating bias fields as Gaussian distributions and maximizing the high-frequency content of the estimated unbiased image. Each image is further normalized to zero mean and unit variance to reduce inter-subject variations. 

Afterwards, to remove large portions of background and reduce model complexity, we utilize Otsu's thresholding \cite{otsu1979threshold} and K-means clustering (k=2) to localize and separate left and right leg-areas. All right legs are mirrored to conform to left legs.

\begin{figure}[tb]
\centering
\includegraphics[width=0.6\linewidth]{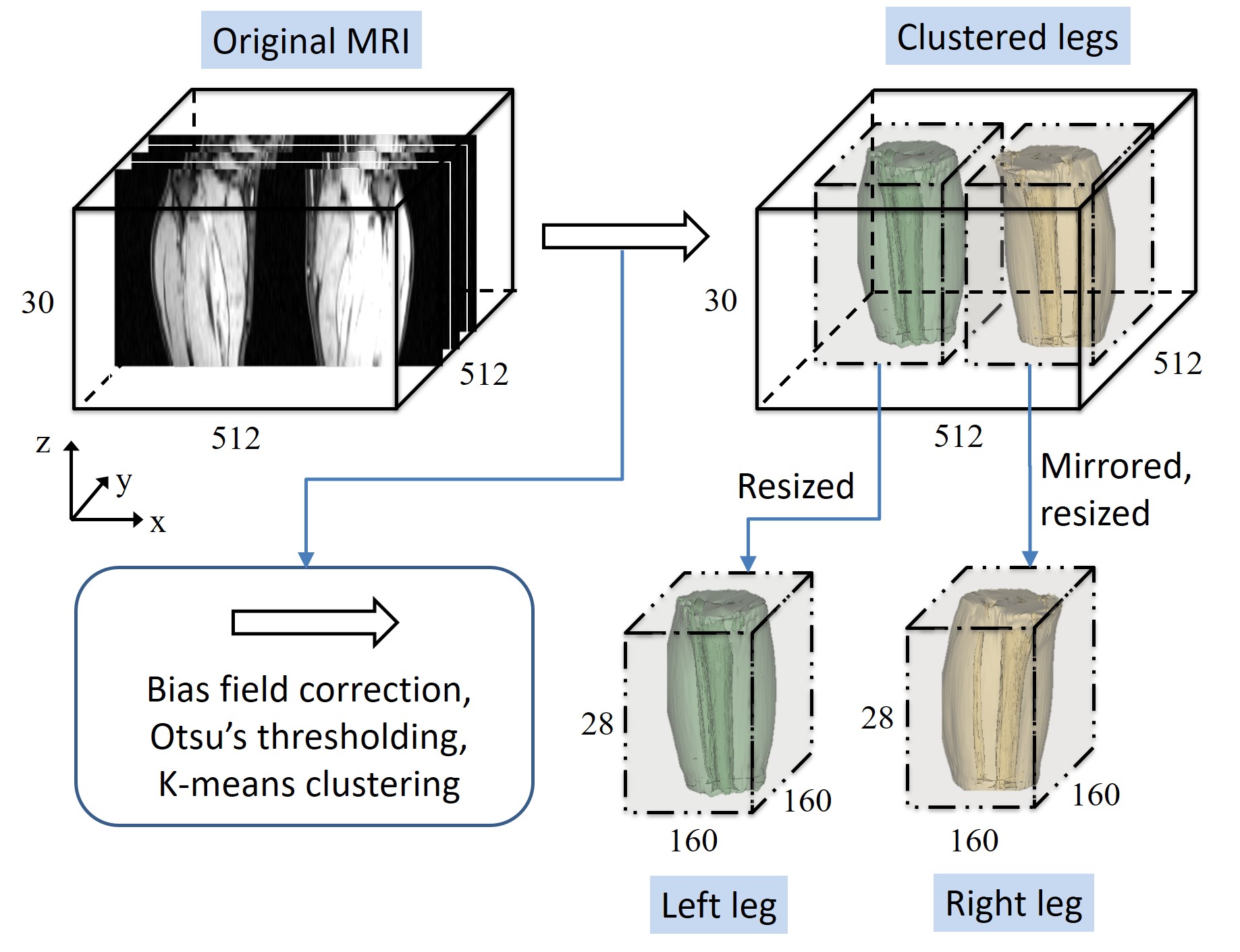}
\caption{Workflow of the pre-processing step. After Bias field correction, Otsu's thresholding, and K-means clustering, each leg-area is localized on the original MR images. The leg-areas are then extracted and resized to get a uniform dimensionality. Right leg is mirrored to left.} \label{fig:pre}
\end{figure}

The workflow and dimensional change of images are shown in Fig.\ \ref{fig:pre}. Note that the pre-processing step is completely unsupervised.

\subsection{FilterNet} 
A typical FCN often consists of an encoder and a decoder. In the encoding phase, multiple levels of features are extracted from the raw input by down-sampling operations to obtain deep and compact representations. The deep representations are then up-sampled to the full resolution in the decoding phase. During the down-sampling and up-sampling steps, a significant problem is the loss of resolution and the associated loss of fine details. UNet \cite{ronneberger2015u, cciccek20163d} and RetinaNet \cite{lin2017focal} include long skip connections to concatenate encoding features to decoding features at the same feature scale to restore  information lost during down-sampling. FC-ResNet \cite{drozdzal2018learning} further uses residual blocks with identity mapping introduced by ResNet \cite{he2016identity} to allow direct gradient back-propagation to earlier layers.

\begin{figure}[tb]
\centering
\includegraphics[width=1\linewidth]{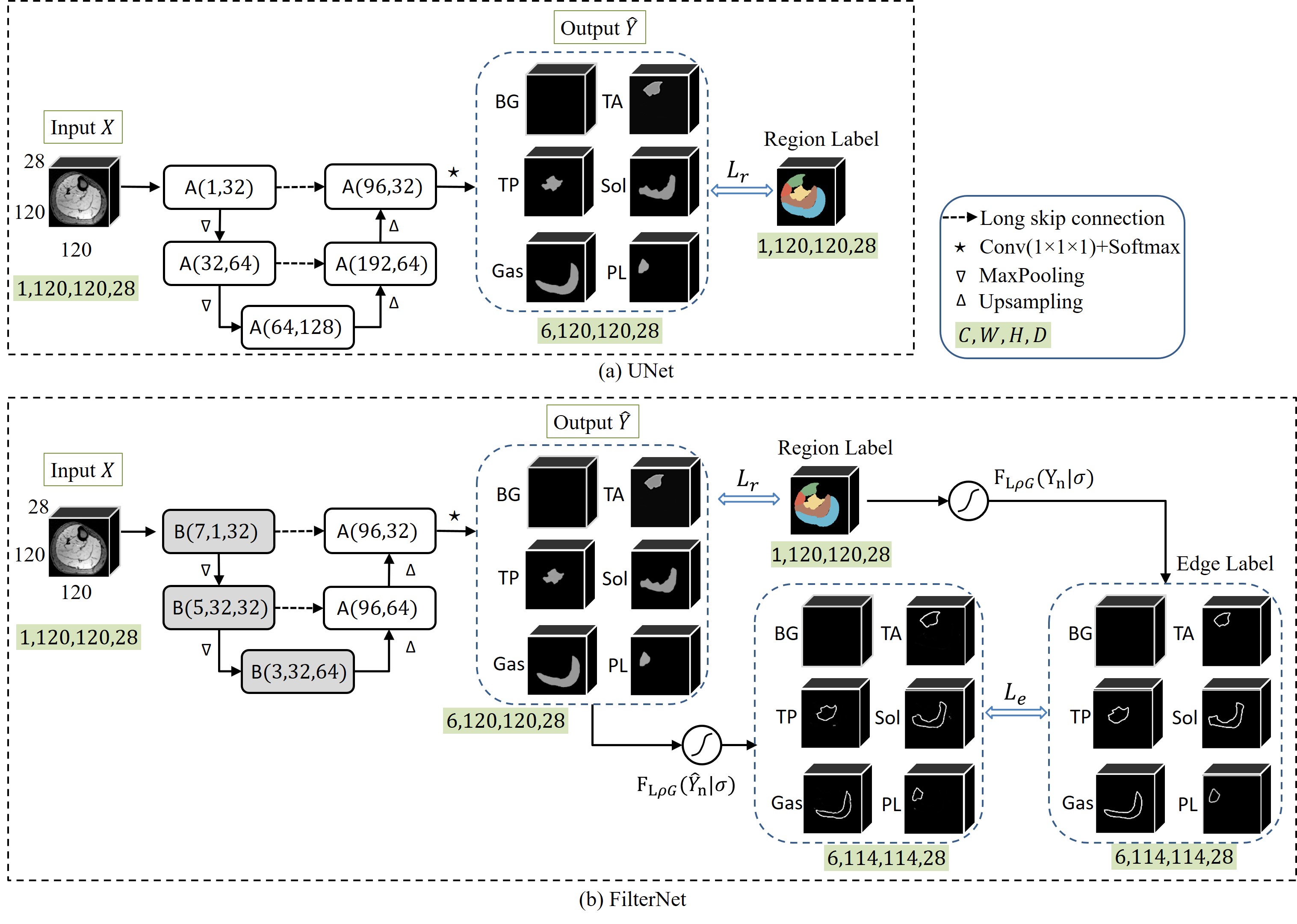}
\caption{UNet (a) vs.\ FilterNet (b) for multi-class calf muscle segmentation. The two networks have the same input and output. [C, W, H, D]: channel, width, height, depth. ``BG'' represents background. Best viewed in color.
} \label{fig:unet}
\end{figure}

Considering the numerous successes that UNet based neural networks achieved in medical image segmentation problems \cite{guo2018deep,amer2019automatic}, we based our FilterNet on a UNet-like architecture (still called UNet for simplicity). The network details of both the base UNet we used and our FilterNet are shown in Fig.\ \ref{fig:unet} and Fig.\ \ref{fig:unet_1}. There are mainly two differences between the two networks. First, in order to preserve fine details and enlarge receptive field during the encoding phase, block B is used in the FilterNet. Block B in the FilterNet utilizes short skip connections to enable gradient identity mapping and therefore preserve fine details. It also provides a portal to increase convolution kernel size that allows the network to take account of a broadened view and utilize contextual information from a large neighborhood. However, a large kernel size significantly increases the number of parameters. To avoid this, FilterNet reduces feature channels in block B at down-sampled scales by a factor of 2 compared with block A used in UNet. Second, the FilterNet further employs an edge gate to extract localized neighborhood relationships in the form of edge detection that allows gradient back-propagation. Different from other works that predict object edges as an extra task to be fused with predictions of object regions \cite{chen2017dcan,shen2017boundary}, FilterNet derives object edge information directly from the object region probability maps through the edge gate and regularizes the derived edge relationship using true edges. In this way, the neighborhood relationship is directly encoded into the region-based probability maps within a single-task FCN, without introducing a number of extra training parameters. In addition, FilterNet implements edge constraints in an end-to-end training manner, it avoids designing and fine-tuning an additional framework for post-processing purpose as \cite{bauer2011fully} and \cite{guo2018deep} did. Details of block A, block B, edge gate, and loss function are described as follows.

\subsubsection{Backbone blocks}
\begin{figure}[tb]
\centering
\includegraphics[width=1\linewidth]{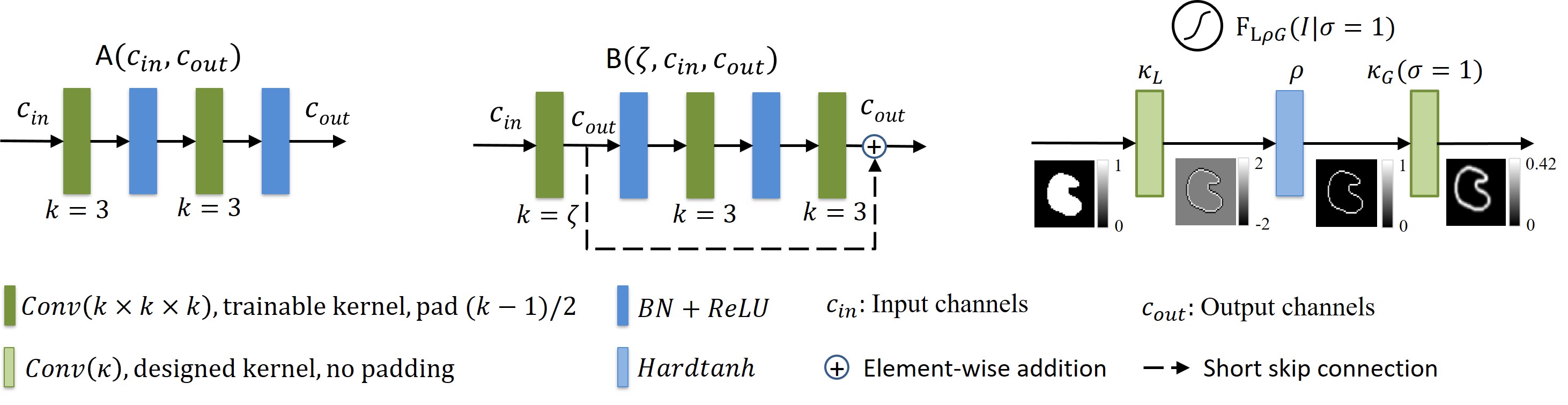}
\caption{Details of block A, block B, and edge gate in FilterNet. An example is shown along with edge gate flow chart. Best viewed in color.
} \label{fig:unet_1}
\end{figure}

Both blocks A and B work as backbone blocks for the FilterNet, as shown in Fig.\ \ref{fig:unet_1}. Block A consists of double layers of convolution ($Conv$) with kernel size $k=3\times 3\times 3$ (padding $p=(k-1)/2$, stride $s=1\times 1\times 1$), batch normalization (BN), and rectified linear unit (ReLU) activation function. Block B adds one more $Conv$ layer with undecided kernel size $\zeta$ at the beginning, and a short skip connection of addition to achieve identity mapping. During the encoding phase, the kernel sizes in block B are set as 7, 5, and 3 for three scales of features to enlarge convolution receptive field. 
As a result \cite{dumoulin2016guide}, FilterNet increases receptive filed size by 22 voxels along each dimension. Note that due to padding and size restriction, the actual increased receptive field size along $z$ axis is less than 22 voxels.

\subsubsection{Edge gate}
Instead of stopping at FCN output of object pixel-wise probability maps as UNet does, an extra step is used in the FilterNet for the edge gate to directly and dynamically derive the true and predicted edge information, respectively, from the ground truth map and the output probability map
and to impose constraints on the predicted edges. For the edge gate to support end-to-end training, kernel-based edge detections are used as convolutions with designed kernels inside the network. In this study, Laplacian of Gaussian (LoG) \cite{sonka2014image} is used. Suppose an input image patch is denoted as $I$, the function of LoG is defined as
\begin{equation}
\label{eq:log}
F_{LoG}(I)=\kappa _{G}*\kappa _{L}*I \; ,
\end{equation}
where $\kappa _{L}$ and $\kappa_ {G}$ represent the Laplacian kernel and the Gaussian smoothing kernel and $*$ is the convolution operation. Since $F_{LoG}$ finds double edges, while the boundary of muscle compartments in this application is not sufficiently clear to define inside and outside edges, we add an activation function to remove the negative edges, and obtain a new function $F_{L\rho G}$. To increase the flexibility of our edge gate and reduce the dependency on handcrafted parameters, the standard deviation $\sigma$ of the Gaussian smoothing function is made trainable. Thus, under the condition of a $\sigma$, the edge gate respond function can be written as  
\begin{equation}
F_{L\rho G}(I|\sigma)=\kappa _{G_{(\sigma)}}*\rho (\kappa _{L}*I) \; ,
\end{equation}
where $\rho $ is a variant of the non-linear hard tanh function that restricts the input value into range $[0, 1]$ such that
\begin{equation}
\label{eq:mask}
\rho (x)=
   \begin{cases}
      0 & x<0 \\
      x & 0\leqslant x\leqslant 1 \\
      1 & x>1
    \end{cases}\; \; .
\end{equation}

The MR images used in this study have in-plane (on $x$-$y$ plane) resolution of 0.7 mm and slice thickness (along $z$ direction) of 7 mm. Therefore, convolution kernels $\kappa _{L}$ and $\kappa _{G_{(\sigma)}}$ are defined as below and applied to $x$-$y$ slices.
\begin{equation}
\label{eq:num}
\kappa _L=\begin{bmatrix}
0 &-1  &0 \\ 
-1 &4  &-1 \\ 
0 &-1  &0 
\end{bmatrix}\in R^{3\times 3}\; ,
\end{equation}
\begin{equation}
\kappa _{G_{(\sigma)}}=\tau [a_{i,j}]_{i,j}\in R^{5\times 5}\; ,
\end{equation}
where $\tau$ is a constant adjusting factor under a given $\sigma$ that ensures $\sum_{i=1}^{5}\sum_{j=1}^{5}a_{i,j}=1$. Element $a_{i,j}$ in $\kappa _{G_{(\sigma)}}$ is defined as,
\begin{equation}
\label{eq:ele}
a_{i,j}=\frac{1}{2\pi \sigma^2}e^{-\frac{(i-\overline{i})^2+(j-\overline{j})^2}{2\sigma^2}}\; .
\end{equation}
In Eq.\ \ref{eq:ele}, $\overline{i}=\overline{j}=3$. 

The gradient back-propagation keeps the fixed kernel $\kappa _{L}$ unchanged and updates kernel $\kappa_ {G}$ with trainable $\sigma$. For the edge gate, an explicit neighborhood relationship is derived from the probability maps.

\subsubsection{Loss function}
For an input image $I\in R^{W\times H\times D}$, the FilterNet predicted output map $\hat{Y}$ can be denoted as,
\begin{equation}
\label{eq:out}
\hat{Y}=D_{A}(E_{B}(I))\in R^{N\times W\times H\times D}\; ,
\end{equation}
where $E_{B}$ is the encoder that uses block B's and $D_{A}$ is the decoder that consists of block A's.

The loss function of FilterNet consists of two terms, $L_{c}$ for region learning amd $L_{e}$ for edge constraining, balanced by a weighting factor $\lambda$ such that
\begin{equation}
\label{eq:loss}
L=(1-\lambda )L_{c}+\lambda L_{e} \; ,
\end{equation}
where $L_{c}$ is a multi-class cross-entropy loss defined as
\begin{equation}
\label{eq:entropy}
L_{c}=\sum_{n=1}^{N}-Y_{n}\cdot log(\hat{Y}_{n}) \; ,
\end{equation}
where $N=6$, $Y_{n}$ is the one-hot encoded region label for class $n$, and $\hat{Y}_{n}\in \hat{Y}$ is the corresponding predicted map for class $n$.
$L_{e}$ represents the least absolute error (L1-norm) loss between the true edge maps and the derived edge maps. $W=\{w_{n}|n=1,...,6\}$ is a weighting array.
\begin{equation}
\label{eq:edge}
L_{e}=\sum_{n=1}^{N}w_{n}\left \| F_{L\rho G}(Y_{n}|\sigma)-F_{L\rho G}(\hat{Y}_{n}|\sigma) \right \|\; .
\end{equation}
During gradient back-propagation, the partial derivative of loss $L$ in terms of $\sigma$ is,
\begin{equation}
\frac{\partial L}{\partial \sigma}=\lambda \frac{\partial L_{e}}{\partial \sigma}\; .
\end{equation}

According to the derivative chain rule, 
\begin{equation}
\frac{\partial L_{e}}{\partial \sigma}=\frac{\partial L_{e}}{\partial \kappa_{G_{(\sigma)}}}\cdot \frac{\partial \kappa_{G_{(\sigma)}}}{\partial \sigma}=\frac{\partial L_{e}}{\partial \kappa_{G_{(\sigma)}}}\cdot [\frac{\partial a_{i,j}}{\partial \sigma}]_{i,j}\; ,
\end{equation}
where $\frac{\partial L_{e}}{\partial \kappa_{G_{(\sigma)}}}$ can be obtained from the differentiation of the Gaussian smoothing $Conv$ layer, and $[\frac{\partial a_{i,j}}{\partial \sigma}]_{i,j}$ is calculated according to the definition of $a_{i,j}$ in Eq.\ \ref{eq:ele}. 

The final multi-class classification output can be defined as the indices of the maximum values on probability maps $\hat{Y}$ along the channel dimension.

FilterNet is optimized by stochastic gradient descent \cite{bottou2010large}. The initial learning rate is $10^{-3}$, which is divided by 5 every 10 epochs. In order to increase the robustness and generalization of the network, the input training patches are sub-regions sized $120\times 120\times 28$, cropped from the localized leg-areas (described in Fig.\ \ref{fig:pre}) with a step size of 20 voxels along x and y directions, which results in 9 times as large the number of the training patches as that of the leg-areas. The batch size is set as 2. We train the network with 30 epochs. Data augmentation is performed, where 3 more patches are generated for each training patch. Namely, a rotation value is randomly chosen between $-10^{o}$ to $10^{o}$, two scaling factors are randomly chosen between 0.8 and 1.2 along x and y directions. The initial value of $\lambda$ is 0.001, multiplied by 10 every 10 epochs. $W$ is $[0, 0.2, 0.2, 0.15, 0.15, 0.3]$ for the 6 classes of background, TA, TP, Sol, Gas, and PL, respectively. The initial value of $\sigma$ is 1.

\section{Experiments and Results}

\subsection{Experimental Setting}
40 lower leg T1-weighted MR images of 40 subjects (10 were healthy, 30 with DM1) were included in this work. The original image size is $512\times 512\times 30$ and the voxel size is 
$0.7\times 0.7\times 7$~mm. The acquisition of these images used the first echo of a 3-point Dixon gradient echo sequence with repetition time (TR) 150~ms, 
echo time (TE) 3.5~ms, field of view (FOV) 36~cm, bandwidth 224~Hz/pixel, and scan time 156~s.
Expert-traced muscle compartment segmentations served as the independent standard. 

Besides FilterNet, we also designed several other neural networks for performance evaluation and comparison purposes. In Section \ref{sec:effective}, based on the differences between UNet and FilterNet, an ablation study was conducted to show the effectiveness of block B and edge gate. Then in Section \ref{sec:comparison}, thorough performance comparisons were presented among the UNet, a multi-task FCN that aggregates region and edge predictions, and FilterNet to demonstrate the superiority and efficiency of the FilterNet. All neural networks were implemented using the PyTorch platform \cite{paszke2017automatic} and applied to the same dataset. The training parameters were identical for these methods. The models were trained on Nvidia GeForce GTX 1070 GPU with 8 GB of memory.

Given a limited-size dataset, 4-fold cross-validation was used to evaluate the performance of each method. The dataset that included both legs was divided in the 4 fold-groups at the subject level so that data from the same subject were never simultaneously used for both training and testing. In 4-fold cross-validation, the 40 subjects were evenly and randomly divided to 4 groups. Each time, one group was taken as the test set, and the remaining three groups were used as the training set. The process was repeated 4 times so every group served as a test set exactly once. As a result, each subject was used for testing just once. DICE Similarity Coefficient (DSC) and absolute surface-to-surface distance (ASSD, in mm) between the automated surface and the manual surface were used as evaluation metrics. For each subject, the performance was averaged for left and right legs.

\subsection{Ablation Study}
\label{sec:effective}
\begin{figure}[tb]
\centering
\includegraphics[width=0.7\linewidth]{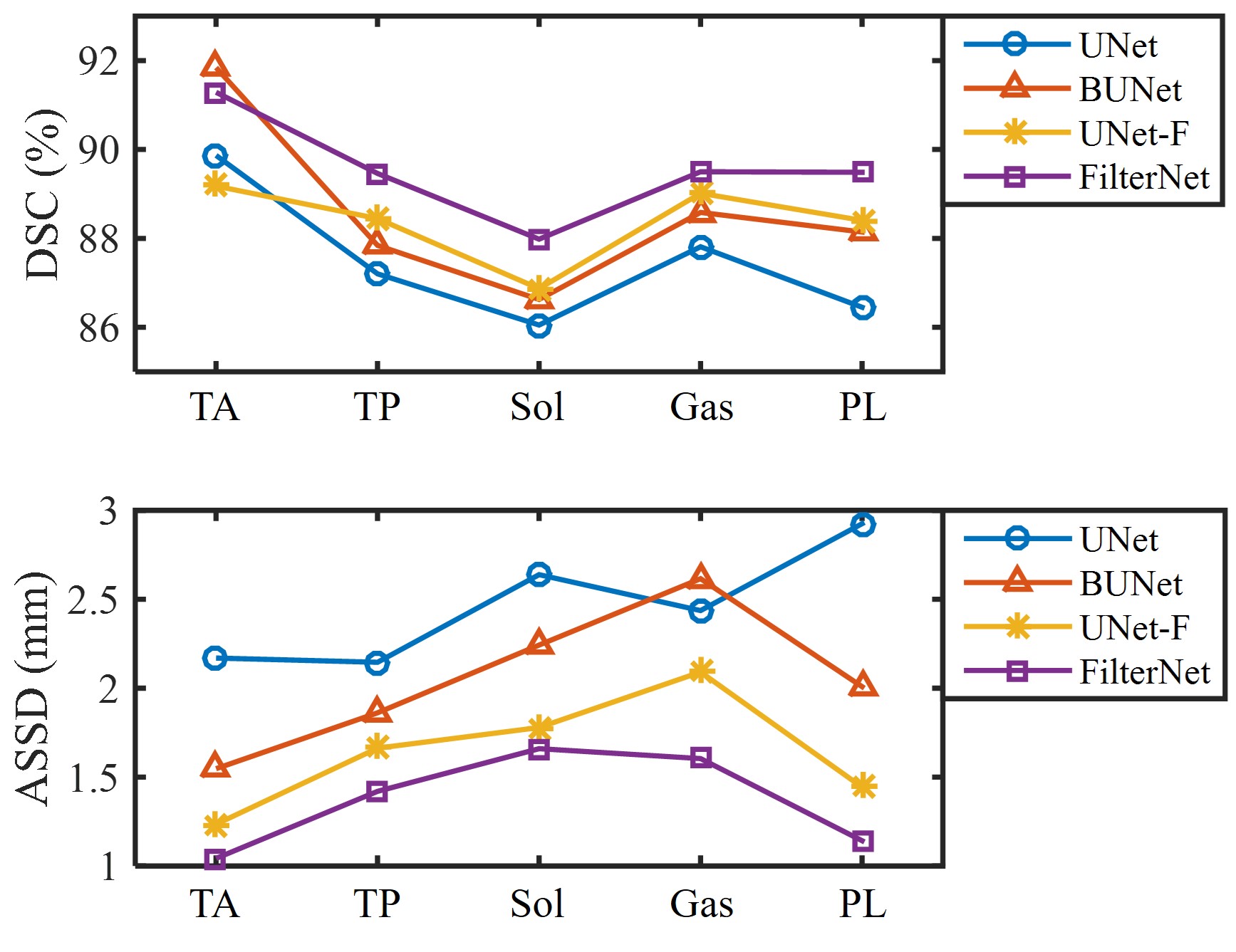}
\caption{Mean values of DICE and ASSD (in mm) from UNet, BUNet, UNet-$F$ and FilterNet. } \label{fig:perfcomp1}
\end{figure}

In order to reveal the performance improvement introduced by block B and edge gate respectively, ablation experiments that included UNet, UNet with block B, UNet with edge gate, and FilterNet were conducted. Therefore, in addition to FilterNet, the other three were described as follows.
\begin{itemize}
\item \textit{UNet:} The details of UNet architecture used in this application are displayed in Fig.\ \ref{fig:unet}(a). UNet output $\hat{Y}$ is
\begin{equation}
\label{eq:outunet}
\hat{Y}=D_{A}(E_{A}(I)) \; ,
\end{equation}
where $E_{A}$ is the encoder consisting of block A's. The loss function is a multi-class cross-entropy loss the same as in Eq.\ \ref{eq:entropy}.
\item \textit{BUNet:} BUNet utilizes block B's in the encoding path of UNet to enlarge convolution receptive field. BUNet output $\hat{Y}$ is the same as FilterNet output in Eq.\ \ref{eq:out}. However, compared with FilterNet, BUNet does not have edge gate. Therefore, $\lambda$ in Eq.\ \ref{eq:loss} is set as 0, such that the loss function here is 
\begin{equation}
L=L_{c}
\end{equation}
\item \textit{UNet-$F$:} Similarly, UNet-$F$ and UNet share the same network architecture. The prediction output $\hat{Y}$ of UNet-$F$ is the same as in Eq.\ \ref{eq:outunet}. However, different from UNet, edge gate is added onto $\hat{Y}$ in UNet-$F$. Thus the loss function of UNet-$F$ is the same as in Eq.\ \ref{eq:loss}.
\end{itemize}

Fig.\ \ref{fig:perfcomp1} shows mean values of DICE and ASSD for five individual muscle compartments obtained from the aforementioned approaches. Overall, FilterNet achieved the best accuracies in terms of DICE and ASSD among the four methods. Both BUNet and UNet-$F$ outperformed UNet. However, UNet-$F$ had better surface positioning accuracy than BUNet. 

\subsection{Performance Comparison}
\label{sec:comparison} 
\begin{figure}[tb]
\centering
\includegraphics[width=1.0\linewidth]{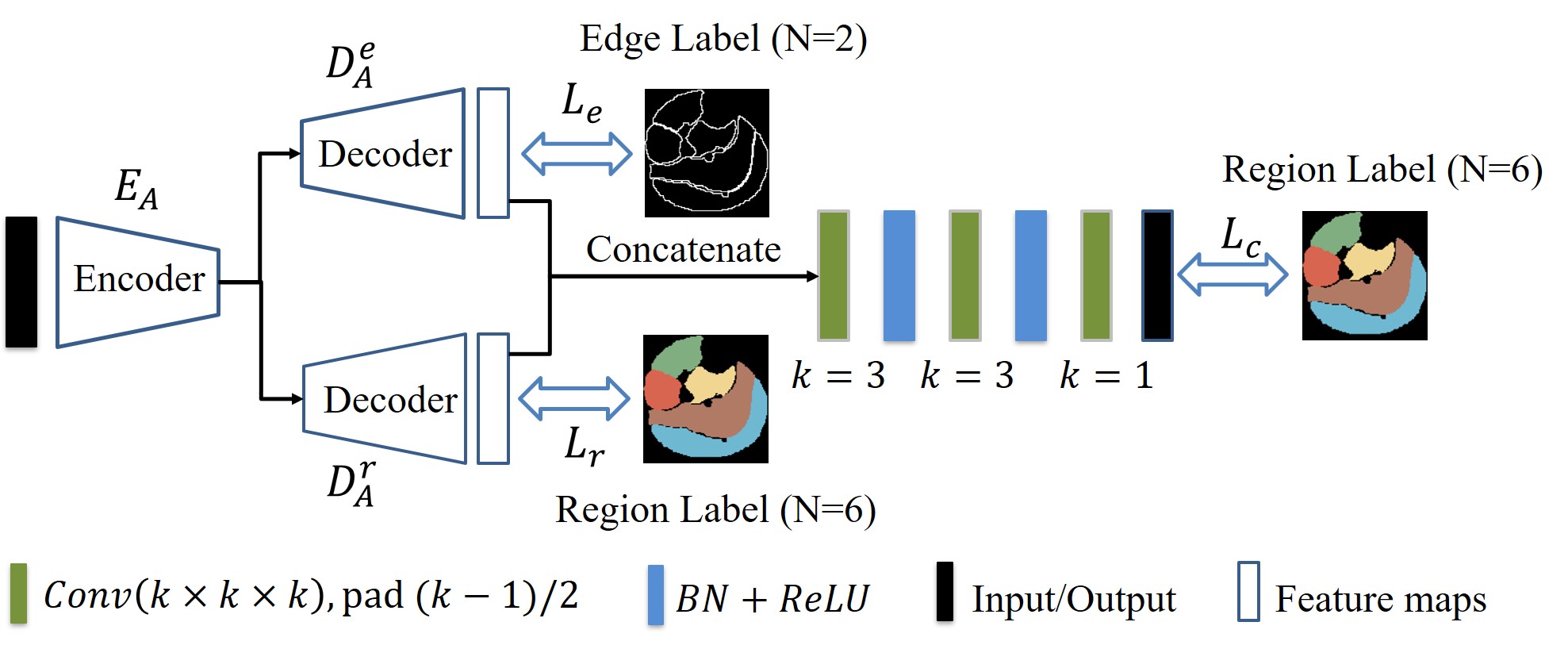}
\caption{The schematic of Boundary-Aware FCN. Best viewed in color.} \label{fig:chen}
\end{figure}

\begin{table*}[tb]
\centering
\def\arraystretch{1}
\caption{DSC and ASSD (mean$\pm$std) for five calf muscle compartments from UNet, Boundary-Aware FCN and FilterNet. The unit for ASSD is mm. Statistical significance in bold.}
\label{tb:result}
\setlength\tabcolsep{1pt}
\begin{tabular}{c|c||c|c|c|c|c|c}
\hline
\multicolumn{2}{c||}{\multirow{2}{*}{Methods}} & \multicolumn{2}{c|}{UNet} & \multicolumn{2}{c|}{Boundary-Aware FCN} & \multicolumn{2}{c}{FilterNet} \\ \cline{3-8}
\multicolumn{2}{c||}{} & Mean$\pm$STD & $p$ value\textsuperscript{*} & Mean$\pm$STD & $p$ value\textsuperscript{*} & Mean$\pm$STD & $p$ value\textsuperscript{*} \\ \hline\hline
\multirow{2}{*}{TA} & DSC & 89.86$\pm$11.07 & \textbf{0.033}  & 90.51$\pm$13.22 & 0.333 &  91.29$\pm$10.11& / \\ \cline{2-8} 
 & ASSD &2.17$\pm$2.00  &\textbf{$\ll$0.001}  & 1.80$\pm$1.97  &\textbf{$\ll$0.001}  & 1.04$\pm$0.81 &/  \\ \hline
\multirow{2}{*}{TP} & DSC &  87.20$\pm$6.89 & \textbf{0.007} &88.11$\pm$6.16  & \textbf{0.043} & 89.46$\pm$4.19 &/  \\ \cline{2-8} 
 & ASSD & 2.15$\pm$1.39  &\textbf{$\ll$0.001}  & 2.10$\pm$1.56 &\textbf{0.002}  & 1.42$\pm$0.66 & / \\ \hline
\multirow{2}{*}{Sol} & DSC & 86.05$\pm$8.94  & \textbf{$\ll$0.001} & 87.09$\pm$8.50 &  \textbf{0.041} &  88.00$\pm$7.92 &/  \\ \cline{2-8} 
 & ASSD & 2.64$\pm$1.87  &\textbf{$\ll$0.001}  & 2.35$\pm$1.87 &\textbf{0.001}  & 1.66$\pm$0.82  &/  \\ \hline
\multirow{2}{*}{Gas} & DSC & 87.81$\pm$10.19  & \textbf{0.017} &88.33$\pm$9.84  & 0.075 &  89.50$\pm$7.99 & / \\ \cline{2-8} 
 & ASSD &2.44$\pm$1.99  &\textbf{$\ll$0.001}  & 2.38$\pm$2.17 &\textbf{0.001}  &  1.60$\pm$1.29 &/  \\ \hline
\multirow{2}{*}{PL} & DSC  &86.45$\pm$12.36  &\textbf{$\ll$0.001}  &89.22$\pm$11.12  & 0.568 & 89.49$\pm$12.44 &/  \\ \cline{2-8} 
 & ASSD &2.93$\pm$3.37  &\textbf{$\ll$0.001}  &1.64$\pm$1.60  &\textbf{0.007}  &1.14$\pm$1.02  &/  \\ \hline
\multicolumn{8}{l}{\textsuperscript{*}\footnotesize{Paired t-test with FilterNet (significance level $p<0.05$)}}
\end{tabular}
\end{table*}
In this section, the performance of FilterNet was thoroughly compared with the performance of UNet and a multi-task FCN, called Boundary-Aware FCN.
\begin{itemize}
\item \textit{Boundary-Aware FCN:} Boundary-Aware FCN follows the basic idea of the kind of FCN proposed in \cite{shen2017boundary}, where the network integrates the predictions for region and edge maps explicitly. The schematic of Boundary-Aware FCN used in this application is shown in Fig.\ \ref{fig:chen}. The input patches, encoder and decoders are the same as those in UNet. One decoder $D_{A}^{r}$ attempts to predict region maps while the other decoder $D_{A}^{e}$ learns the corresponding edge maps. Then the predicted region and edge maps are concatenated and fed into several $Conv$, $BN$, and $ReLU$ layers (denoted as $\Phi$) to get the final region prediction $\hat{Y}_{c}$. Thus, $\hat{Y}_{c}$ can be described as,
\begin{equation}
\hat{Y}_{c}=\Phi(\hat{Y}_{r}\odot \hat{Y}_{e}) \; ,
\end{equation}
where $\odot$ is the concatenation operation, $\hat{Y}_{r}=D_{A}^{r}(E_{A}(I))$ is the predicted region map, and $\hat{Y}_{e}=D_{A}^{e}(E_{A}(I))$ is the predicted edge map. 

The loss function $L$ includes three cross-entropy loss terms, region loss $L_{r}$, edge loss $L_{e}$, and a final combined loss $L_{c}$.
\begin{equation}
L=\sum_{i=e,r,c}L_{i}=\sum_{i=e,r,c}-Y_{i}log(\hat{Y}_{i}) \; ,
\end{equation}
where each $Y_{i}$ is the corresponding label map. Note that our implementation of Boundary-Aware FCN uses the same encoder and decoder as described in our baseline UNet design, which is different from the original description in \cite{shen2017boundary}.
\end{itemize}

As a result, Table \ref{tb:result} summarizes DSC and ASSD between the automated segmentations and the independent standard on five calf muscle compartments for the UNet, Boundary-Aware FCN and FilterNet. Compared with UNet, FilterNet generated significantly better results for each compartment in terms of both DICE and ASSD. FilterNet was also shown to have significant differences from Boundary-Aware FCN in DICE for TP and Sol and in ASSD for each muscle compartment. 

\begin{figure}[t]
\centering
\includegraphics[width=1.0\linewidth]{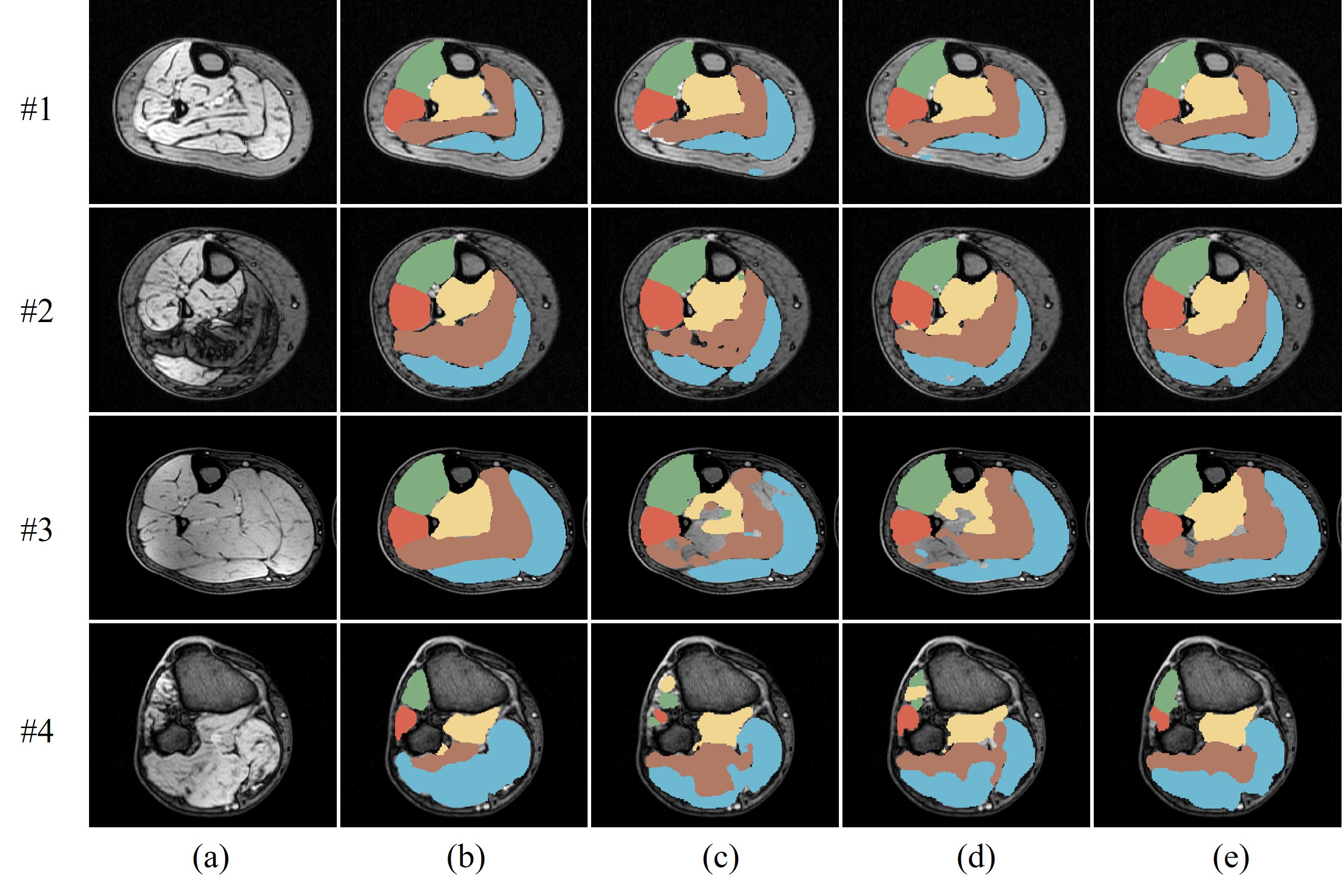}
\caption{Cross-sectional segmentation examples overlaid with MR images. (a) Original scan. (b) Ground truth. (c) UNet. (d) Boundary-Aware FCN. (e) FilterNet. Each row represents a 2D cross-sectional example from a different image. Best viewed in color.} \label{fig:result1}
\end{figure}

From top to bottom, Fig.\ \ref{fig:result1} displays four 2D segmentation examples from images of four patients, with each representing a unique situation. The first example shows a normal subject with calf muscle surrounded by a thick SAT layer. In this case, Gas segmented by UNet had leakage into Sol and SAT, while Gas and Sol segmented by Boundary-Aware FCN also leaked into the SAT layer. The second example is from an MR image of a patient with severe DM1, where TA segmentation obtained from UNet spread into TP and PL, as well as holes existing in Sol. TP segmentation from Boundary-Aware FCN had false positives in true PL, and a hole appeared in Gas. The third example shows notable intensity inhomogeneity around the boundary of TP and Sol. Both UNet and Boundary-Aware FCN were sensitive to the inhomogeneity. The last example presents a 2D cross-sectional slice that is near to one end of the lower leg, where the muscle compartment boundaries are complicated and tough to identify. UNet and Boundary-Aware FCN generated strangely shaped Sol and some voxels inside the true TA were misclassified as TP. In contrast to the situations happened to UNet and  Boundary-Aware FCN, though the segmentations from FilterNet were not always perfect, FilterNet was able to relief these problems and appeared to be more robust to image inhomogeneity and object shape maintenance on these examples.

\begin{figure}[t]
\centering
\includegraphics[width=1\linewidth]{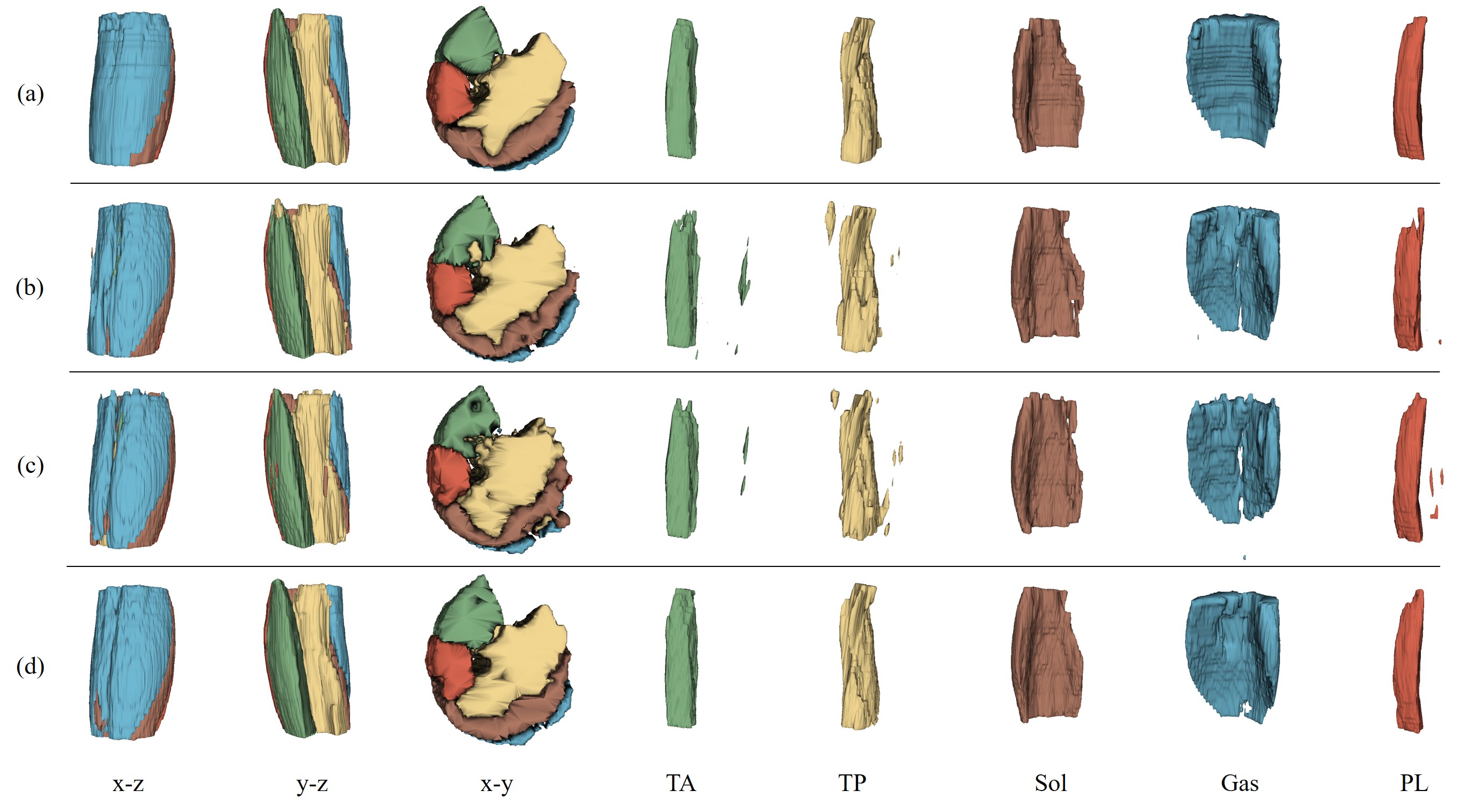}
\caption{3D demonstration of muscle compartment segmentations from the same leg. Each row represents a method while each column represents a view category. (a) Ground truth. (b) UNet. (c) Boundary-Aware FCN. (d) FilterNet. The first three columns are three orthogonal views of the five muscle compartment segmentations, followed by five columns showing individual segmentations. Note the extra objects generated by UNet and Boundary-Aware FCN for TA, TP, Gas and PL. Best viewed in color.
} \label{fig:result2}
\end{figure}

Fig.\ \ref{fig:result2} lists the 3D shapes of the muscle segmentations from the same leg of a patient. From the $x$-$y$ view, FilterNet generated smoother and topologically superior segmentations. When observing the individual muscle compartment models, UNet and Boundary-Aware FCN showed obvious region leakages of TA, TP, and PL, while FilterNet segmentations were free from such leakages.
Table \ref{tb:comp} compared the number of model parameters, memory usage and averaged training time per epoch. FilterNet has the lowest number of parameters and memory usage, and UNet ran the fastest during training.

\begin{table}[tb]
\centering
\def\arraystretch{1}
\caption{Number of model parameters, memory usage, and averaged training time per epoch for UNet, Boundary-Aware FCN and FilterNet. Best performance in bold.}
\label{tb:comp}
\setlength\tabcolsep{1pt}
\begin{tabular}{c||c|c|c}
\hline
Methods & \begin{tabular}[c]{@{}c@{}}\# parameters\\ (Millions)\end{tabular} & \begin{tabular}[c]{@{}c@{}}Memory\\ (GB)\end{tabular} & \begin{tabular}[c]{@{}c@{}}Training time \\(mins)\end{tabular} \\ \hline\hline
UNet   &1.97 & 4.67 & \textbf{11} \\ \hline
Boundary-Aware FCN  &  2.39&  6.01& 25 \\ \hline
FilterNet &  \textbf{1.44}&  \textbf{3.76}&  16\\ \hline
\end{tabular}
\end{table}

\section{Discussion}
\subsection{Ablation Study}
The superiority of BUNet against UNet as shown in Fig.\ \ref{fig:perfcomp1} indicates the effectiveness of block B. 
However, the calculated ASSD index for Gas segmentation by BUNet became worse compared with UNet. Gas has a long and flexible shape on the $x$-$y$ plane and also shares similar appearance patterns with nearby objects, when neighborhood was broadened while proper edge-aware regulations were lacking, false positives might have been increased across the whole scope and reflected in worsened surface positioning accuracy. 

When edge gate was added, UNet-$F$ generated segmentations with apparent improvement in ASSD over UNet. The loss term associated with edge gate in Eq.\ \ref{eq:edge} implies that a) when a pixel is not close to an object boundary, either inside or outside of the object, the probability values of all voxels located inside the neighborhood that belongs to the same class should be similar while b) when a pixel is close to an object boundary, probability value differences between such voxels may show dissimilarity. Therefore, with edge gate, UNet-$F$ was able to impose neighborhood-wise constraints to the predicted output and achieved more topologically correct results.

Compared with UNet, FilterNet reached a new level of accuracy. As we mentioned earlier, when neighborhood is broadened, edge-aware regulations are needed to reduce false positives and strengthen boundary information. Likewise, when edge-aware regulations are presented, broadened neighborhood helps reduce sensitivity to local noise and results in improved segmentation accuracy. Therefore, the two mechanisms can benefit from each other and lead to the best performance of FilterNet in the ablation study.

\subsection{FilterNet vs.\ UNet, Boundary-Aware FCN}
UNet is a single-task FCN that includes only one decoder to learn object region maps, while Boundary-Aware FCN is a multi-task FCN with two decoders to learn region and edge maps respectively and extra layers to fuse the output of the two tasks, with the cost of substantially increased numbers of parameters and computational time, as shown in Table \ref{tb:comp}. With emphasized boundary information and integration of region and edge learning, Boundary-Aware FCN achieved superior performance in terms of DICE and ASSD compared with UNet (Table \ref{tb:result}).  

Both FilterNet and Boundary-Aware FCN intended to take advantage of object boundary to improve object segmentation accuracy. Boundary-Aware FCN learns object boundary from scratch and regards predicted boundary maps as extra feature channels into region learning. While FilterNet derives object boundary directly from the predicted region maps and encodes the neighborhood-wise relationship by designed edge detectors with little cost. Instead of learning from scratch, FilterNet avoided introducing a lot of extra parameters to learn the desired edge pattern. The edge detectors turned out to be more efficient than an edge-learning decoder, from the numerical and visual comparisons in Table \ref{tb:result}, Fig.\ \ref{fig:result1}, and Fig.\ \ref{fig:result2}. There are two potential reasons behind such phenomena. First, given a small training dataset, a model with a huge amount of parameters may result in over-fitting, and thus the performance may be compromised. Second, calf muscle compartment boundaries are hard to learn, due to the fact that all muscle compartments share similar appearance patterns. Besides, edge-like appearances caused by intramuscular nerves and vessels \cite{fuxe1965distribution,sinha2006vivo} can be misleading. 

Fig.\ \ref{fig:result1} and Fig.\ \ref{fig:result2} further demonstrated that when multiple objects are close to each other and share very similar textures, optimization that is based on pure pixel-level classification loss may cause disjoint regions or holes inside the true object. When neighborhood-wise dissimilarity penalty were added for voxels away from the boundary, as edge gate did in FilterNet, the situations of disjoint regions and holes were mitigated. 

\subsection{Approach and Future Work}
The application of calf muscle compartment segmentation represents a category of multi-class segmentation problems, where nearby objects are next to each other and have very similar textures. In order to segment each object accurately and at the same time maintain object shape topology, we proposed FilterNet as a neighborhood relationship enhanced FCN that has broadened convolution receptive field and an edge detector based gate to apply constraints directly to the probability maps in an end-to-end training manner. 

Besides increasing convolution kernel size as we did with block B, there are other ways to broaden the receptive field or integrate context information from a larger scope. For example, dilated convolutions \cite{yu2015multi} can be applied to enlarged ranges without increasing filter sizes. Attention \cite{chen2016attention} has the potential to take into account of multi-scale features simultaneously by trainable weights. Exploration of more efficient ways to enlarge feature neighborhood will remain as future work. 

For the edge gate used in our FilterNet, Laplacian edge detector and Gaussian smoothing kernel with trainable $\sigma $ were applied to derive object edges. After optimization, we obtained a $\sigma $ of 0.89, 0.85, 0.92 and 0.90 pixel for each fold, respectively. 
We have also explored a sole use of the Laplacian edge detector in the edge gate, $F_{L\rho G}$ with fixed $\sigma$ ($\sigma=1$) while leaving largest connected component for each label of the results to be generated by $F_{L\rho G}(\sigma=1)$ as a post processing step. Note that only keeping the largest connected component is only feasible when the desired object is known to be a single region while our method does not have this restriction. It turns out that though the performance differences in terms of DSC are small, the performance differences in terms of ASSD are notable. As the ASSD values shown in Table \ref{tb:last}, our current FilterNet outperforms the other tested methods. This means that the usage of Gaussian smoothing makes the edge gate more robust to noise and the neural network has the ability to fine-tune hyper-parameters like $\sigma$ to further improve the performance. Finding advanced convolution-based detectors with trainable hyper-parameters is also worth exploring in the future.
\begin{table}[tb]
\centering
\def\arraystretch{1}
\caption{Averaged ASSD for each muscle compartment. $F_L$: FilterNet with only Laplacian kernel used in the edge gate. $F_{L\rho G}(\sigma=1)$: FilterNet with fixed $\sigma$=1 in the Gaussian kernel. $F_{L\rho G}(\sigma=1)$ \& LLC: leaving largest connected component for each label is applied to the results of $F_{L\rho G}(\sigma=1)$. $F_{L\rho G}(\cdot |\sigma)$: FilterNet with trainable $\sigma$. Best performance in bold.}
\label{tb:last}
\setlength\tabcolsep{1pt}
\begin{tabular}{c||c|c|c|c|c}
\hline
Method & TA & TP & Sol & Gas & PL \\ \hline
$F_L$ & 1.57 & 1.85 & 2.28 & 2.24 & 2.02 \\ \hline
$F_{L\rho G}(\sigma=1)$ & 1.48 & 1.84 & 2.11 & 2.15 & 1.85 \\ \hline
$F_{L\rho G}(\sigma=1)$ \& LLC & \textbf{1.01} & 1.56 & 1.69 & 1.65 & 1.21 \\ \hline
$F_{L\rho G}(\cdot |\sigma)$ & 1.04 & \textbf{1.42} & \textbf{1.66} & \textbf{1.60} & \textbf{1.14} \\ \hline
\end{tabular}
\end{table}
In addition to reducing the sensitivity to noise, Gaussian smoothing also increases edge response area sizes, given extremely small portion of edge voxels in a volume. Due to  the sparsity of edge response areas and the optimization efficiency of L1-norm in this case \cite{melkumova2017comparing}, the L1-norm was used in Eq.\ \ref{eq:edge} instead of the L2-norm.  

Use of edge constraints in a light-weight manner was previously considered, e.g., by Ronneberger \textit{et al.} \cite{ronneberger2015u} who computed a weight map that highlighted the edge areas and calculated the weighted cross-entropy loss. This approach increases the importance of edge areas, but the weight map calculation is based on ground truth only. Our edge gate is applied to prediction maps during each back-propagation, thus has a high penalty in holes and undesired disjoint objects that may otherwise appear during prediction. In \cite{al2018spnet}, deep features are used to directly learn shape parameters that reflect shape differences from a mean shape representation for cervical vertebrae on X-ray images. In this way, disjoint regions are eliminated and the edge is smooth. Cervical vertebrae are rigid with stable (position-independent) shapes, while muscle compartments consist of soft tissues that may change their shapes dramatically in response to changes in  position, large deviations from the mean shape representation result. A boundary loss term based on the summation of nearest distances from segmentation pixels to the true boundary in level set representation was proposed in \cite{kervadec2019boundary} with pixels given penalties according to the distance from the true boundary. However, since the calculation is only carried out for pixels segmented as the target, regions corresponding to holes inside the prediction do not have any penalty associated with them. As an highly relevant improvement, our method penalizes both holes and regions away from the true edges while being computationally efficient. In this study, we applied edge constraints on the axial plane only, due to the extreme anisotropic resolution of the dataset. The clinical reality of MR imaging protocols that are used when imaging calf muscles unfortunately results in such anisotropic acquisition parameters. If more isometric data become available, we plan to use 3D edge detectors in the future and expect to see additional performance gains.

Different from muscle segmentation by atlas-based methods applied to MR images from homogeneous populations \cite{nguyen2019robustly}, the dataset of 40 images we used in this study was quite diverse and included unaffected patients (10/40) and patients with DM1 (30/40). For patients with DM1, the posterior compartment muscles, gastrocnemius and soleus, are usually the earliest and eventually most severely affected muscles. The changes are not limited to these muscles. However, for the purpose of measuring changes over time for eventual clinical trials, being able to track muscles that are less severely affected may be equal or more important than those that are affected earliest and most severely in DM1. In this work, we have shown the feasibility and robustness of our approach in segmenting individual calf muscle compartments in normals and when DM1 was presented. In the future, for each muscle compartment segmentation, disease-affected regions can be easily clustered from the healthy muscle structure to develop compartment-based biomarkers guiding a disease progression study in the clinic.

\section{Conclusion}
A neighborhood relationship enhanced FCN was reported and applied to individual calf muscle compartment segmentation on T1-weighted MR images. With an increased convolution receptive field, resolution-preserving skip connections, and explicitly edge-aware regulations by a kernel-base edge gate to constrain pixel-level probability values inside a neighborhood, our FilterNet considered the specialty of adjacent multi-class muscle segmentation and delivered a striking performance improvement (DSC$>$0.88, ASSD$<$1.66~mm) over previously-reported methods (DSC$<$0.55, ASSD$\sim$7~mm), suggesting clinical-use feasibility of automated calf-compartment segmentation.

\section*{Acknowledgment}
We would like to thank Eric Axelson for preparing the data and Ashley Cochran for performing the manual tracings. 

This work was supported in part by the NIH grants R01-EB004640 and R01-NS094387.

\section*{References}

\bibliography{mybibfile}

\end{document}